\documentclass[prb,preprint]{revtex4-1} 
\usepackage{amsmath}
\usepackage{amsfonts}
\usepackage{graphicx}

\begin{document}
\title{Apollo's Voyage: A New Take on Dynamics in Rotating Frames}
\date{\today}

\author{Ujan Chakraborty}
\email{uc15ms074@iiserkol.ac.in, ujanchakraborty2@gmail.com}
\author{Ananda Dasgupta}
\email{adg@iiserkol.ac.in}
\affiliation{Department of Physical Sciences, \\ Indian Institute of Science Education and Research Kolkata}

\begin{abstract}
We first demonstrate how our general intuition of pseudoforces has to navigate around several pitfalls in rotating frames. And then, we proceed to develop an intuitive understanding of the different components of the pseudoforces in most general accelerating (rotating and translating) frames: we show that it is not just a sum of the contributions coming from translation and rotation separately, but there is yet another component that is a more complicated combination of the two. Finally, we demonstrate using a simple example, how these dynamical equations can be used in such frames.

\textit{The following article has been submitted to the American Journal of Physics. After it is published, it will be found at Link.\cite{Link}}
\end{abstract}

\maketitle

\section{Introduction}
The diurnal motion of the Sun across the sky from east to west has
perhaps been the most significant and most universal natural occurence
in the lives of human beings since antiquity. There was no lack of
theories among our ancients, on the Sun's apparent motion across the
firmament during the day, and absence at night. The ancient Egyptians
pictured the Sun god Ra sailing his mighty ship across the ethereal
waters over the earth at day, and in the underworld at night,\cite{Ra} while
the Greeks believed that Apollo had been tasked with pulling the enchained
Sun across the skies everyday, on his golden chariot drawn by four
mighty stallions.\cite{Apollo} The east did not lack theories either: in ancient
Hindu mythology, it was Surya riding across the heavens in his chariot
drawn by the seven fiery steeds, with Aruna as his charioteer,\cite{Surya}
while for the Shinto of Japan, it was the goddess Amaterasu coming
out of her divine palace to preside over the mortals.\cite{Amaterasu} It was not until
the early Renaissance, that the heliocentric model of the solar system
was founded: \begin{quote}
	``Nicolaus Copernicus Thorunensis, terrae motor, solis caelique stator''
\end{quote} (``Nicolaus Copernicus, son of Toru\'{n}, mover of the earth, stopper
of the sun and heavens'').\cite{Nicolaus Copernicus} Here, we show that a reflection on the apparent diurnal motion of the Sun might help modify some of our intuition regarding dynamics in accelerating frames.

\section{An Apparent Paradox}
\subsection{A Problem with our Intuition}
We begin with a problem, and an observation that goes against the
common intuition regarding ``pseudoforces''. We observe from the Earth, that the Sun appears to move around the
Earth once in about 24 hours, but know that actually the Earth rotates
on its axis once in the same time. The Earth also revolves around the Sun in about $365\frac{1}{4}$
days, but this motion due to the gravitational attraction between
the two would be ignored at all points in our discussion for simplicity.
One can see directly that even its inclusion would not affect our
conclusion qualitatively, and a back-of-envelope calculation reveals
that the quantitative changes are negligible, so this approximation
really isn't remotely similar to the spherical chicken approximation,
but rather akin to considering the spherical earth (rather than the
oblate spheroid earth) approximation. Of course, the entire observation can be explained quite simply
from a space-fixed (inertial) frame. But, suppose an observer on the
rotating Earth (a non-inertial frame) decides to address the situation
using ``pseudoforces''. This should be possible without much ado
using the tools one learns in a basic undergraduate course on Newtonian
mechanics. So, the following question was posed to quite a large number
of students who had taken such a course in the same level as the textbook by Kleppner and Kolenkow,\cite{Kleppner Kolenkow} and also to some students who had also taken a more advanced course on the same, based on the celebrated textbook by Goldstein.\cite{Goldstein} For such an observer (someone
on the Earth), which force on the Sun causes it to revolve around
the Earth? A common answer received was, the ``centripetal'' force.
This answer is technically correct, but not at all enlightening. It is a bit like the saying: ``If you know when a tree was planted,
you can determine its age quite accurately''. Because, ``centripetal'' is a name assigned to just any force directed
centrally inwards: in other words, a force that is responsible for
the centripetal acceleration is the centripetal force. It is the dynamical
origin of such a force that we are interested in here: how it arises
through physical interactions, or as a pseudoforce. However, in spite
of being formally trained in the basic physics of non-inertial frames,
quite a large number of students failed to correctly identify the
origin of this ``centripetal'' force. The gravitational attraction between
the Earth and the Sun is not really relevant in this particular problem,
since we do not take consider effects due to the rotation of the Earth
round the Sun. So, the only forces relevant in this scenario are the
pseudoforces. While accounting for the forces
of relevance in this scenario, most students successfully identified only the ``centrifugal''
(pseudo)force. This is perhaps because it is common intuition built
up from day-to-day experience (e.g. when a moving vehicle brakes suddenly),
that for an observer in an accelerated frame, the pseudoforce on an
object is equal in magnitude to the mass of the object multiplied
by the acceleration of the frame, and directed opposite to the acceleration
of the frame. And this recipe yields the centrifugal force in this
case. However, observe that though the centrifugal force is equal
in magnitude to the required ``centripetal'' force, the direction
is opposite. The centrifugal force on the Sun is directed radially
outward, so it cannot support the observation of the Sun moving around
the Earth once in a day, just by itself. It appears that we have here
a paradox: one has an object rotating around the observer, but, cannot
identify the dynamical origin of the centripetal force (the force
that provides the centripetal acceleration)! 

Another example, of a different physical setting, but of essentially
the same nature, where we encounter the same ``paradox'': an observer sitting at the centre of a rotating merry-go-round (a carousel). Perhaps this might appear more appropriate to some readers, since we do not have to make approximations like neglecting the revolution of the Earth around the Sun. For simplicity, let us assume that the origin of the merry-go-round-fixed-frame
is a point on its axis of rotation, and that the angular velocity
is constant. Our observer shall find that the surroundings execute circular motion
around him or her. The centrifugal force would be directed outwards
from the axis. Along the same lines as in the previous paragraph,
one might think that this is the only force of interest here (neglecting gravity and all other such forces, which are orders of
magnitude less). So, the acceleration of the surrounding bodies can be expected to
be horizontal and radially outward from the centre of the merry-go-round.
But this is evidently false: in the rotating frame, the surroundings
revolving around the axis must have an acceleration directed towards
the axis itself, and not outwards. So, who provides the centripetal
force?

The answer to both problems, is that the pseudoforce in a rotating
frame does not constitute of the centrifugal force alone. There is
yet another force which we usually ignore in most day-to-day scenarios.
It is the ``Coriolis force''. It does not occur to us immediately,
perhaps because in most observations to which we apply our intuition
of pseudoforces, Coriolis forces do not affect the dynamics significantly
over short time scales. For example, in most cases, students imagine
Coriolis forces as those which deflect winds and ocean currents, or
decide which way a whirlpool is going to swirl on different sides
of the Equator. It usually does not cross one's mind that the same
Coriolis forces might be crucially relevant for observations as simple
as the diurnal motion of the Sun. In the discussion below, we see
how the Coriolis force resolves the situation completely, and try
to build an intuition about the true nature of this force (there is
good reason why this intuition is often lacking).

\subsection{The Resolution}

As a first step towards a resolution, we look at the following result
from Goldstein.\cite{Goldstein} When we take the derivative of vector $\boldsymbol{G}$
with respect to time in a rotating frame, we have, 
\begin{equation}
\left(\frac{d\boldsymbol{G}}{dt}\right)_{s}=\left(\frac{d\boldsymbol{G}}{dt}\right)_{r}+\boldsymbol{\omega}\times\boldsymbol{G}\label{rotation formula}
\end{equation}
where $s$ stands for the space-fixed frame, and $r$ for the frame
rotating with angular velocity $\boldsymbol{\omega}$. We see that
the time derivative in a rotating frame, $\left(\frac{d\boldsymbol{G}}{dt}\right)_{r}$,
is different from its counterpart in a non-rotating frame, $\left(\frac{d\boldsymbol{G}}{dt}\right)_{s}$.
This can be rationalized in the following fashion: the unit vectors
in a rotating frame, unlike their translating counterparts, are not
constant. Hence, once has to account for that in the derivative. One
also sees that for any translating but non-rotating frame, $\boldsymbol{\omega}=\boldsymbol{0}$,
and hence, $\left(\frac{d\boldsymbol{G}}{dt}\right)_{translating}=\left(\frac{d\boldsymbol{G}}{dt}\right)_{s}$.
Thus, instead of an inertial frame, $s$ can be any non-rotating frame
having arbitrary translational motion for the above Eq. (\ref{rotation formula})
to be satisfied.

We address both the merry-go-round problem and the Earth's rotation
problem simultaneously, as the difference between the two is essentially
superficial.\cite{in the end it doesn't even matter} Consider two frames, with origins coinciding on the axis of the
merry-go-round (or, the axis of the Earth's rotation, which means that our observer is positioned on the North or the South
Pole, or that we have chosen to ignore the radius of the Earth, an
omission the reader will undoubtedly pardon). One is $s$, non-rotating, the other is $r$, fixed to the merry-go-round
(or the Earth), and rotating with it. Then, let $\boldsymbol{r}$
be the coordinate of some the particle whose motion is under study
(or the Sun, which we treat here as a particle). In both frames $r$
and $s$, the coordinate $\boldsymbol{r}$ reads the same, since the
origin is the same. Furthermore, let us assume that $\boldsymbol{\omega}$
is constant. Then, taking the derivative of $\boldsymbol{r}$ twice,
each time making adjustments by Eq. (\ref{rotation formula}), we get the
following equation governing the motion in the rotating frame $r$:

\begin{equation}
\boldsymbol{F}-2m\left(\boldsymbol{\omega}\times\boldsymbol{v_{r}}\right)-m\boldsymbol{\omega}\times\left(\boldsymbol{\omega}\times\boldsymbol{r}\right)=m\boldsymbol{a_{r}}\label{force equation in uniformly rotating frame}
\end{equation}

where $\boldsymbol{v_{r}}$ and $\boldsymbol{a_{r}}$ represent the
particle's velocity and acceleration respectively in the rotating
frame, and we have made the substituion $\boldsymbol{F}=m\left(\frac{d^{2}\boldsymbol{r}}{dt^{2}}\right)_{s}$,
by Newton's laws. A detailed calculation is shown in the appendix
(\ref{subsec: Appendix force equation in uniformly rotating frame}).

Now, if $\boldsymbol{F}_{pseudo}$ be the pseudoforces in the rotating
frame $r$, we should demand $\boldsymbol{F}+\boldsymbol{F}_{pseudo}=m\boldsymbol{a_{r}}$.
Hence, we have: $\boldsymbol{F}_{pseudo}=\boldsymbol{F_{co}}+\boldsymbol{F_{ce}}$.
The second term is the all-too-familiar centrifugal term, $\boldsymbol{F_{ce}}=-m\boldsymbol{\omega}\times\left(\boldsymbol{\omega}\times\boldsymbol{r}\right)$.
The first term is the one of greater interest here (even if only because
it often escapes our intuition): $\boldsymbol{F_{co}}=-2m\left(\boldsymbol{\omega}\times\boldsymbol{v_{r}}\right)$,
the Coriolis term. The Coriolis term is the one responsible for counter-intuitive
behaviour. For the `surroundings', stationary in the $s$ frame, $\boldsymbol{v_{r}}=-\boldsymbol{\omega}\times\boldsymbol{r}$,
and hence, the Coriolis term equals twice of the centrifugal term
with a reversed sign. This provides a resolution of the problem.

Thus, we see that the folklore of pseudoforces being equal to the
mass of the particle times the negative of the frame's acceleration
does not really hold true in rotating frames. Why does the Coriolis
term not appeal to our intuition as much as the `negative of frame's
acceleration term'? After all, all of us are in a rotating frame (on
the surface of the Earth)! The answer lies in the fact that the Coriolis
force is proportional to the cross product of the angular velocity
of the frame $\boldsymbol{\omega}$, and the\textit{ velocity of the
observed particle} in the frame, $\boldsymbol{v_{r}}$. For an observer
on the surface of the Earth, $\boldsymbol{\omega}$ is approximately
$7.3\times10^{-5}s^{-1}$. So, except for bodies moving at very, very
high $\boldsymbol{v_{r}}$, its effects are practically negligible
over short intervals of time. Of course, over long intervals of time,
even relatively small $\boldsymbol{v_{r}}$'s do cause significant
deviations, depending on the angle between $\boldsymbol{\omega}$
and $\boldsymbol{v_{r}}$, e.g. for ocean currents and winds. But
for objects moving at sufficiently high $\boldsymbol{v_{r}}$, the
Coriolis term is definitely not neglible, compared to the centrifugal
term. In fact, as demonstrated, for objects fixed respect to that
frame, with respect to which our concerned frame is rotating at angular
velocity $\boldsymbol{\omega}$, the Coriolis term is twice the centrifugal
term with an opposite sign, and their resultant provides the centripetal.
This is exactly why an observer on the surface of the Earth observes
the Sun to orbit the Earth diurnally, in spite of the centrifugal
term directed radially outward.

\section{Dynamical Equations, in the Most General Accelerating Frame}

Having discussed some of the kinematic consequences of an observer
being in a rotating frame, we now turn to the dynamics in a most general
accelerating frame: a frame that is rotating (no longer at a constant
$\boldsymbol{\omega}$) as well as translating. One may think, shouldn't
the dynamics in such a frame simply be the sum of the effects due
to translation without rotation, and due to rotation without translation?
Though naively this appears to be a plausible solution, we shall see
that this is not entirely correct. The force equation concerned can
still be visualized as a combination of the effects due to translation
and rotation, but it is slightly more involved than just their sum.
We demonstrate below, that, in the pseudoforce, considering the contributions
due to the rotation in absence of translation ($\boldsymbol{F_{rot}=}\boldsymbol{F_{ang}}+\boldsymbol{F_{co}}+\boldsymbol{F_{ce}}$)
and the contribution due to translation in absence of translation
($\boldsymbol{F_{trans}}=-m\boldsymbol{A_{r}}$), we still need to
add yet another term $\boldsymbol{F_{comb}}$, which is zero if the
frame is either stationary (coinciding with $s$) or non-rotating.

In what follows, $s$ will denote a stationary
/ inertial frame, while $r$ will be an accelerating frame.

\subsection{The Force Equations, in the Most General Case}

We now can no longer assume that the origins of the $s$ frame and
the $r$ frame coincide. Let $\boldsymbol{r_{s}}$ and $\boldsymbol{r_{r}}$
be the coordinates of the particle under study in the $s$ (inertial)
and $r$ (non-inertial) frames respectively. Let \textbf{$\boldsymbol{R}$}
the (time-dependent) vector between the origins, that is, $\boldsymbol{r_{s}}=\boldsymbol{r_{r}}+\boldsymbol{R}$.
Then, $\boldsymbol{R}$ and its derivatives $\boldsymbol{V_{r}}=\left(\frac{d\boldsymbol{R}}{dt}\right)_{r}$
and $\boldsymbol{A_{r}}=\left(\frac{d^{2}\boldsymbol{R}}{dt^{2}}\right)_{r}$
capture the translational dependence.

Let $\boldsymbol{\omega_{r}}$ be the angular velocity of the frame,
$\left(\frac{d\boldsymbol{\omega_{r}}}{dt}\right)_{r}=\left(\frac{d\boldsymbol{\omega_{r}}}{dt}\right)_{s}=\boldsymbol{\dot{\omega}_{r}}$.

We have, 
\begin{equation}
\boldsymbol{F}_{pseudo}=\boldsymbol{F_{trans}}+\boldsymbol{F_{rot}}+\boldsymbol{F_{comb}}
\end{equation}
where
\begin{align}
	\boldsymbol{F_{trans}} &= -m\boldsymbol{A_{r}} \\
	\boldsymbol{F_{rot}} &= \boldsymbol{F_{ang}}+\boldsymbol{F_{co}}+ \boldsymbol{F_{ce}}
\end{align}
with
\begin{align}
	\boldsymbol{F_{ang}} &=- m\boldsymbol{\dot{\omega}_{r}}\times\boldsymbol{r_{r}} \\
	\boldsymbol{F_{co}} &= -2m\left(\boldsymbol{\omega_{r}}\times\boldsymbol{v_{r}}\right) \\
	\boldsymbol{F_{ce}} &= -m\boldsymbol{\omega_{r}}\times\left(\boldsymbol{\omega_{r}}\times\boldsymbol{r_{r}}\right)
\end{align}
and 
\begin{equation}
\boldsymbol{F_{comb}}=-2m\boldsymbol{\omega_{r}}\times\boldsymbol{V_{r}}-m\boldsymbol{\omega_{r}}\times\left(\boldsymbol{\omega_{r}}\times\boldsymbol{R}\right)-m\boldsymbol{\dot{\omega}_{r}}\times\boldsymbol{R}
\end{equation}
Then,
\begin{equation}
\boldsymbol{F}+\boldsymbol{F}_{pseudo}=m\boldsymbol{a_{r}}\label{Newton's laws in most general accelerating frame}
\end{equation}

The derivations are in the appendix (\ref{subsec: Appendix Newton's laws in most general accelerating frame}).

\subsection{Considering a Few Special Cases}

The last stated Eq. (\ref{Newton's laws in most general accelerating frame})
may be considered to be the working form of Newton's law in a most
general accelerating frame. All this might appear a bit cumbersome.
It is easier to analyze a few simple cases:
\begin{enumerate}
\item There is a complete agreement with our common intuition when the frame
is non-rotating ($\boldsymbol{\omega}=\boldsymbol{0}=\boldsymbol{\dot{\omega}}$):
$\boldsymbol{F}_{pseudo}=-m\boldsymbol{A_{r}}$.
\item When it is rotating uniformly ($\boldsymbol{\dot{\omega}_{r}}=\boldsymbol{0}$)
but not translating, it is reasonable to set $\boldsymbol{R}=\boldsymbol{0}$,
and then $\boldsymbol{F}_{pseudo}=\boldsymbol{F_{ce}}+\boldsymbol{F_{co}}$.
We see that this too is in complete agreement, with Eq. (\ref{force equation in uniformly rotating frame}). 
\item When it is rotating (perhaps non-uniformly) but not translating, again,
it is reasonable to set $\boldsymbol{R}=\boldsymbol{0}$, and then
we have $\boldsymbol{F}_{pseudo}=\boldsymbol{F_{ang}}+\boldsymbol{F_{co}}+\boldsymbol{F_{ce}}$.
That is, we have the minimal modification to Eq. (\ref{force equation in uniformly rotating frame}),
simply by addition of the term $\boldsymbol{F_{ang}}$, just as we
might expect.
\item When there is both translation and rotation, the situation is more
involved. However, it might be interesting to pose the following questions:
given $\boldsymbol{R}$, $\boldsymbol{V_{r}}$, does there exist $\boldsymbol{\omega_{r}}$,
$\boldsymbol{\dot{\omega}_{r}}$ such that $\boldsymbol{F_{comb}}=\boldsymbol{0}$,
and vice versa? Perhaps this question is similar in spirit somewhat
to Chasles' theorem.
\end{enumerate}
We provide an interesting and instructive example below to demonstrate how
these equations might be used.

\subsection{An Example: the Merry-Go-Round Archer}

Suppose an archer decides to try out shooting an arrow from a merry-go-round.
What additional factors should be taken into account?

The first thing to observe, is that once the arrow leaves the bow (literally),
there is no relevant force acting on it (we are only interested in the horizontal motion). Gravity does not affect the horizontal motion of the arrow, except
for limiting the range, so we leave that consideration to the archer's
usual skill. The vertical motion of the arrow (due to gravity / the
angle decided upon by the archer) is not affected by the rotation,
as $\boldsymbol{\omega}$ has a non-zero component only in the vertical
direction. Of course, we also ignore the drag force due to the air,
and assume the arrow doesn't go ``whichever way the wind doth blow''. So a stationary observer should observe the arrow to fly off in
a straight line (we ignore the vertical motion of the arrow
). But what
does the archer see?

Suppose, we consider the archer to be sitting on the rotating merry-go-round
at a distance of $r_{0}$ from its axis of rotation. And, the
merry-go-round is rotating on its axis at a constant angular speed
$\omega$, anticlockwise. We consider the inertial frame $s$
to have origin on the axis of the merry-go-round, and for the non-inertial
frame $r$ (the archer) to be using cylindrical coordinates $\left(\boldsymbol{\hat{\rho}},\boldsymbol{\hat{\theta}},\boldsymbol{\hat{z}}\right)$
where $\boldsymbol{\hat{\rho}}$ is directed towards the axis of the
merry-go-round, and $\boldsymbol{\hat{z}}$ is directed upwards. So,
$\boldsymbol{R}=-r_{0}\boldsymbol{\hat{\rho}}$, $\boldsymbol{\omega_{r}}=\omega\boldsymbol{\hat{z}}$,
$\boldsymbol{V_{r}}=\boldsymbol{0}$, and $\boldsymbol{A_{r}}=\boldsymbol{0}$.
Here, $\boldsymbol{F_{comb}}=-m\boldsymbol{\omega_{r}}\times\left(\boldsymbol{\omega_{r}}\times\boldsymbol{R}\right)=-m\omega^{2}r_{0}\boldsymbol{\hat{\rho}}$,
$\boldsymbol{F_{ang}}=\boldsymbol{F_{trans}}=\boldsymbol{0}$, $\boldsymbol{F_{co}}=2m\omega\boldsymbol{v_{r}}\times\boldsymbol{\hat{z}}$,
$\boldsymbol{F_{ce}}=m\omega^{2}\boldsymbol{r_{r}}$. Hence, we have
\begin{equation}
\boldsymbol{a_{r}}=\omega^{2}\left(\boldsymbol{r_{r}}-r_{0}\boldsymbol{\hat{\rho}}\right)+2\omega\boldsymbol{v_{r}}\times\boldsymbol{\hat{z}}
\end{equation}

Of course, one may solve this second order differential equation with
suitable initial conditions to get the full trajectory. However, a
qualitative analysis of what happens just at the instant the arrow
is shot, is quite easy. At that instant, $\boldsymbol{r_{r}}=\boldsymbol{0}$,
and so $\boldsymbol{a_{r}}\left|_{t=0}\right.=-\omega^{2}r_{0}\boldsymbol{\hat{\rho}}+2\omega\boldsymbol{v_{r}}\left|_{t=0}\right.\times\boldsymbol{\hat{z}}$.
$-\omega^{2}r_{0}\boldsymbol{\hat{\rho}}$ looks the acceleration
due to a centrifugal (pseudo)force, but it actually follows from \textbf{$\boldsymbol{F_{comb}}$}.
When $r_{0}=0$ (that is, the archer is at the center of the merry-go-round),
the arrow is observed to be simply deflected to the right by the Coriolis
term $2\omega\boldsymbol{v_{r}}\times\boldsymbol{\hat{z}}$, as expected.

\section{Conclusion}

We have demonstrated some of the limitations of our common intuition when dealing with accelerating frames, in particular, failure in rotating frames. That being done, we have provided a natural way to extend our intuition to such scenarios. In this article, we have restricted ourselves to dynamics of point particles, that is, equations involving force and acceleration only. It is interesting to note that similar studies can be carried out for systems of particles, in particular, in the setting of rigid body dynamics. One may study the relations between torque, angular momenta, and angular velocities from general accelerating frames as well, in the same spirit as of this article. We are presently compiling such an article.\cite{article2} This leads us to obtain an alternate proof (and more importantly, a new physical realization) of Euler's equations for rigid body motion, in the same lines as in Mott's 1966 deduction.\cite{Mott1966}

\appendix

\section{Deduction of the Dynamical Equation in a Frame Rotating at a Constant
Angular Velocity:\label{subsec: Appendix force equation in uniformly rotating frame}}

From Eq. (\ref{rotation formula}), with $\boldsymbol{r}$ as $\boldsymbol{G}$:

\begin{equation}
\left(\frac{d\boldsymbol{r}}{dt}\right)_{s}=\left(\frac{d\boldsymbol{r}}{dt}\right)_{r}+\boldsymbol{\omega}\times\boldsymbol{r}
\end{equation}

And again, with $\left(\frac{d\boldsymbol{r}}{dt}\right)_{s}$ as
$\boldsymbol{G}$:

\begin{equation}
\left(\frac{d^{2}\boldsymbol{r}}{dt^{2}}\right)_{s}=\left(\frac{d^{2}\boldsymbol{r}}{dt^{2}}\right)_{r}+2\boldsymbol{\omega}\times\left(\frac{d\boldsymbol{r}}{dt}\right)_{r}+\boldsymbol{\omega}\times\left(\boldsymbol{\omega}\times\boldsymbol{r}\right)
\end{equation}

\section{Deduction of the Dynamical Equation in a Most General Accelerating
Frame:\label{subsec: Appendix Newton's laws in most general accelerating frame}}

For a translating-cum-rotating frame, a straightforward calculation
by considering the time derivative of $\boldsymbol{r_{s}}$ in the
two frames by Eq. (\ref{rotation formula}) yields:
\begin{equation}
\boldsymbol{v_{s}}=\left(\boldsymbol{v_{r}}+\boldsymbol{V_{r}}\right)+\boldsymbol{\omega_r}\times\left(\boldsymbol{r_{r}}+\boldsymbol{R}\right)\label{eq:velocity formula}
\end{equation}
\begin{equation}
\boldsymbol{a_{s}}=\left(\boldsymbol{a_{r}}+\boldsymbol{A_{r}}\right)+2\boldsymbol{\omega_{r}}\times\left(\boldsymbol{v_{r}}+\boldsymbol{V_{r}}\right)+\boldsymbol{\omega_{r}}\times\left(\boldsymbol{\omega_{r}}\times\left(\boldsymbol{r_{r}}+\boldsymbol{R}\right)\right)+\boldsymbol{\dot{\omega}_{r}}\times\left(\boldsymbol{r_{r}}+\boldsymbol{R}\right)
\end{equation}

Using $m\boldsymbol{a_{s}}=\boldsymbol{F}$, this can be written as:
\begin{align}
m\boldsymbol{a_{r}} & =\boldsymbol{F}-m\boldsymbol{A_{r}}-2m\left(\boldsymbol{\omega_{r}}\times\boldsymbol{v_{r}}\right)-m\boldsymbol{\omega_{r}}\times\left(\boldsymbol{\omega_{r}}\times\boldsymbol{r_{r}}\right)-m\boldsymbol{\dot{\omega}_{r}}\times\boldsymbol{r_{r}}\nonumber \\
 & -2m\left(\boldsymbol{\omega_{r}}\times\boldsymbol{V_{r}}\right)-m\boldsymbol{\omega_{r}}\times\left(\boldsymbol{\omega_{r}}\times\boldsymbol{R}\right)-m\boldsymbol{\dot{\omega}_{r}}\times\boldsymbol{R}
\end{align}

We can immediately identify the second term on the right hand side as arising out of the translational motion of the frame, the third through fifth terms as arising out of the rotational motion of the frame, and the last three terms as a combination of both translation and rotation.

\begin{acknowledgements}
	UC would like to thank the INSPIRE scholarship (2015) and the KVPY scholarship (2016-2020) for support during his degree programme.
\end{acknowledgements}


\begin{thebibliography}{99}
\bibitem{in the end it doesn't even matter} It might appear that the merry-go-round problem is a bit easier to
	address in detail compared to the problem of the Sun's diurnal motion, since one does not have the additional complication of the translational motion of the frame (one less detail to wrap one's head around). But it is not so: as additional translational motions do not contribute	anything to the picture, ``in the end it doesn't even matter''.
\end{thebibliography}
\end{document}